\documentstyle[12pt,epsfig,rotate,aasms4]{article}

\newcommand\be{\begin{equation}}
\newcommand\ee{\end{equation}}

\newcommand{\mcommand}[1]{\ifmmode #1\else $#1$\fi}
\newcommand\Rsat{\mcommand{R_{\rm sat}}}
\newcommand\Urect{\mcommand{U_{\rm rect}}}
\newcommand\Umask{\mcommand{U_{\rm mask}}}
\newcommand\xsource{\mcommand{x_{\rm s}}}
\newcommand\ysource{\mcommand{y_{\rm s}}}
\newcommand\fresnelC{\mcommand{{\cal C}}}
\newcommand\fresnelS{\mcommand{{\cal S}}}
\def\sci#1{\mcommand{\times 10^{#1}}}

\def\unit#1{\mcommand{\;{\rm #1}}}
\newcommand{\kelvin}{\unit{K}}
\newcommand\meter{\unit{m}}
\newcommand\second{\unit{s}}
\newcommand\as{\unit{as}}
\newcommand\marcsec{\unit{mas}}
\newcommand\mas{\marcsec}
\newcommand\parsec{\unit{pc}}
\newcommand\kpc{\unit{kpc}}
\newcommand\yr{\unit{yr}}
\newcommand\km{\unit{km}}
\newcommand\kmps{\unit{km\;s^{-1}}}
\def\micron{\unit{\mu m}}
\newcommand\kg{\unit{kg}}
\newcommand\Day{\unit{day}}
\newcommand\AU{\unit{AU}}
\newcommand\angstrom{\unit{\AA}}

\def\vec#1{\mbox{\boldmath$#1$\unboldmath}}

\newcommand\etal{et al.}
\newcommand\etc{etc.}

\newcommand\ltsim{\lesssim}
\newcommand\gtsim{\gtrsim}

\begin{document}

\rightline{\bf CWRU-P7-97}
\rightline{July 1997}
\rightline{Submitted to: \em Publications of the Astronomical Society of the
Pacific}

\title{The Improved Resolution and Image Separation (IRIS) Satellite}
\author{Glenn D. Starkman}
\affil{Department of Physics and Department of Astronomy, Case Western Reserve
University, Cleveland, OH 44106-7079\\
Electronic mail: {\tt gds6@po.cwru.edu}}
\and
\author{Craig J. Copi}
\affil{Department of Physics, Case Western Reserve
University, Cleveland, OH 44106-7079 \\
Electronic mail: {\tt cjc5@po.cwru.edu} \\
IRIS home page: {\tt http://erebus.phys.cwru.edu/$\sim$iris}}
\authoraddr{10900 Euclid Avenue\\ Cleveland, OH 44106-7079}

\begin{abstract}
Natural (such as lunar) occultations have long been used to study sources on
small angular scales, while coronographs have been used to study high contrast
sources.  We propose launching the Improved Resolution and Image
Selection~(IRIS) Satellite, a large steerable occulting satellite.  IRIS will
have several advantages over standard occulting bodies.  IRIS blocks up to
$99.97\%$ of the visible light from the occulted point source.  Because the
occultation occurs outside both the telescope and the atmosphere, seeing and
telescope imperfections do not degrade this performance.  If placed in Earth
orbit, integration times of $250$--$2500\second$ can be achieved over 95\% of
the sky, from most major telescope sites.  Alternatively, combining IRIS with a
basic 2-m (or 4-m) space telescope at the Earth-Sun L2 point could yield
longer integration times and simplified orbital dynamics. Applications for
IRIS include searching for planets around nearby stars, resolution of
micro-lensed LMC and Galactic bulge stars into distinct image pairs, and
detailed mapping of solar system planets from the ground.  Using ground-based
K-band observations, a Jovian planet with a surface temperature of
$400\kelvin$ or Venus at $700\kelvin$ could be imaged at all angular
separations greater than about $0.35\as$ out to over $10\parsec$.  Space based
observations in the K-band (in emission) or B-band (in reflection) could do
better.  Resolution of microlensed stars would greatly improve our
understanding of the Massive Compact Halo Objects (MACHOs) responsible for
observed microlensing events and comprising 20-90\% of the mass of our galaxy.
\end{abstract} 


\section{Introduction}

The search for planets around nearby stars is a major objective of modern
astronomy.  Recently, astronomers claimed to have discovered Jupiter-mass
planets orbiting nearby stars (see e.g., Mayor \& Queloz~1995; Butler \&
Marcy~1996; Mayor \& Queloz~1995; Cochran, \etal~1997; Marcy, \etal~1997;
Noyes, \etal~1997).  They observed the periodic variation in the central
stars' velocities due to their motion around the center-of-mass of the
star-planet systems.  However, they were unable to image the planets
directly---the planets are too close to the stars that they orbit, so
diffraction in the telescope and atmospheric seeing cause the starlight to
wash out the planets.  This problem is generic to observations of compact,
compound sources in which the range of brightnesses of the individual
components is large.

On another forefront, many astrophysical systems have structure on
milli-arcsecond or sub-milli-arcsecond angular scales. For example, when a
massive object passes sufficiently close to the line-of-site to a star, the
image of the star splits into multiple images, with the sum of the
brightnesses of all the observable images exceeding the brightness of the
unlensed star (so-called ``micro-lensing'').  This
gravitational amplification has recently been detected (see e.g., Beaulieu,
\etal~1995; Paczynski, \etal~1995; Alcock \etal~1996) for stars in both the
bulge of our Galaxy and in the nearby Large Magellanic Cloud.  It is caused by
massive compact objects (MACHOs) of approximately $0.1$--$1.0$ solar masses in
the Galactic halo.  MACHOs apparently comprise 20--90\% of the mass of the
Galaxy, yet little is known about them.  This is partly because the individual
images are $\lesssim 1\marcsec$ apart and cannot be resolved by existing
ground or space-based telescopes.

Many systems have interesting features on milli-arcsecond scale.  Binary stars,
globular cluster cores, Galactic supernovae, the accretion disk around a
massive black hole at the center of the Milky Way, cores of nearby galaxies,
lensed images of galaxies, and distant galaxies and clusters all have
structure on the milli-arcsecond scale. Apparent angular displacements due to
parallax and proper motions are also on this angular scale for objects in the
nearest few kpc of the Galaxy.  The proper motion of a star at $10\kpc$, with
a velocity relative to the Sun of $200\kmps$, is approximately $5\mas\yr^{-1}$.

The separation of dim sources from nearby bright ones, and the resolution of
images at milli-arcsecond scales in the UV, optical, and near IR is limited
principally by two effects: atmospheric seeing and diffraction.  Seeing
constrains the angular resolution to $\Delta\theta \ge \theta_{\rm seeing}
\gtsim 0.2\as$ from the ground.  Diffraction forces
$\Delta\theta\geq\theta_{\rm diff}\simeq 1.2\lambda/D$.  Here $\lambda$ is the
wavelength of the observations and $D$ is the diameter of the telescope or the
baseline of the interferometer.

The effects of seeing on resolution can be reduced using adaptive optics (AO)
to nearly the diffractive limit in the IR, not yet at shorter wavelengths.
However, the diffractive limit at $1\micron$ on a 10-m telescope is still
$25\mas$, far too large to resolve structures like Galactic microlenses.
Moreover, while AO is typically very good at squeezing the central core of the
point spread function (PSF) to near the diffraction limit, it still leaves the
halo of the PSF\@.  Getting all the way to the brightness contrasts of interest
for planet searches ($>10^7$) strictly on the basis of AO, is considered
technologically challenging in the IR, and daunting at shorter wavelengths.

Seeing can be eliminated entirely by observing from space (though scattering
inside the telescope persists), however, even then diffraction persists.
Greater resolution can be obtained using interferometry, and plans exist to do
both ground-based and space-based IR and optical interferometry.  A
space-based interferometer the same size as IRIS is intrinsically superior to
IRIS for high resolution imaging, though likely more costly, and more
challenging technologically.  For high contrast imaging, the benefits of being
diffraction limited are also considerable, with reduced light pollution at
large angular separations from a bright source.  However, diffraction patterns
do fall relatively slowly with angular separation, making it more difficult to
achieve very high contrasts.  Interferometers are also not immune to problems
of light scattering in the telescope.

The starting point for such another approach transits the sky each night.
Lunar occultations have long been used to resolve small-scale angular
structure (see e.g., Han \& Gould~1996; Simon \etal~1996; Mason~1996; Richichi
\etal~1996; Cernicharo \etal~1994; Adams \etal~1988).  As the Moon orbits the
Earth, it sweeps eastward across the night sky.  If it occults another source,
by monitoring the light-curve from that source, one can deduce the integrals
of the surface-brightness of the source perpendicular to the apparent lunar
velocity as a function of angular position parallel to the lunar velocity.

There are however several problems with using the Moon in this fashion:
(1) The lunar orbit is fixed; sources cannot be scheduled for occultation.  
Some sources are never occulted.
The Moon is unlikely to occult the same object more than once a decade, 
so it is impossible to build up statistics, repeat an observation,
or recover from poor observing conditions.
(2) The apparent angular velocity of the Moon 
with respect to the background stars ($0.55\as\second^{-1}$) is 
large and fixed by the orbital speed of the Moon.
If a source is too dim to obtain good photon statistics during an occultation,
there is no way to reduce the apparent velocity 
nor  to repeat the observations to improve the statistics.
This makes it difficult to use the Moon to study dim sources such
as planets. 
(3) The Moon is bright despite its low albedo.
Even the dark side is visible to the naked eye.

To overcome or reduce these difficulties, we are investigating the scientific
merits of launching a large occulting satellite.  The Improved Resolution and
Image Separation (IRIS) Satellite will consist of a large opaque patch or
a series of patches.  Patch geometry will be optimized for both high-resolution
image reconstruction and separation of bright from dim sources.  The simplest
configuration would be a single square patch which travels parallel to one
pair of sides for separation of dim images from bright ones (such as planet
detection), and is rotated through $\pi/4$ (so that the leading and trailing
edges are not parallel to each other) for image reconstruction at ultra-high
resolution.  Different patch geometries will undoubtedly prove optimal for
different types of observations.

There are two possible configurations in which such a satellite could be
deployed.  The first is a square $200$--$300\meter$ on a side in a high apogee
eccentric Earth orbit using ground-based telescopes for all observations.
Such a program has advantages in cost and risk since no astronomical
instruments are deployed in space.  Moreover, the satellite could be used from
any telescope, including those of amateur astronomers, if satellite
coordinates were made publicly available.  We refer to this as the
ground-space configuration.  A major complication is likely to be the
difficulty of ensuring adequate stability and predictability of the orbital
dynamics given magnetospheric and atmospheric drag, and gravity gradients.
None of these issues has yet been fully addressed.

Alternately, the satellite could be deployed at a Lagrangian point of the
Earth-Sun system (most likely L2) in conjunction with a simple 2-m (or 4-m)
class space telescope.  This would have the disadvantage of requiring a space
telescope, however, orbital dynamics are likely to be more stable,
predictable, and adjustable.  Furthermore, the absence of atmospheric seeing
will allow a smaller occulter to be used.  We refer to this as the space-space
configuration.  We will concentrate below on the ground-space configuration
because our investigations of it are more advanced, with greater attention to
the space-space configuration in future publications.

Given its size, one needs to do whatever possible to minimize IRIS's mass.
One approach is to fashion IRIS out of thin opaque balloons inflated with low
pressure gas.  Problems of reflected Earth and Sun-light could be addressed by
shielding.  Additionally, the underside could be reflective so that only
photons from a known field would be reflected into the telescope on Earth.
Careful choice of the orientation of the satellite relative to the line of
sight, could ensure that this field was sufficiently dark.  A combination of
both approaches may be needed to contain the problems of reflected Sun, Earth
and Moon-light.  Alternately, one could allow (or even excite) 
small oscillations of the earth-facing surface
and thus smear the reflected fields over the full field-of-view.
The problem of light reflection would be mitigated somewhat in the space-space
configuration where the Earth and Moon are dimmer and subtend smaller angles.
Power demands could be met by solar cells covering as much of the satellite's
surface as is necessary.

Especially in the space-ground configuration,
gross adjustments to the orbit to enable scheduling of desired
targets will require either ion rockets or other alternative drive
mechanisms discussed below.
No astronomical instruments will be on board the satellite,
significantly reducing costs and improving reliability.  Furthermore in the
space-ground configuration any improvements in telescope or detector design
can be immediately used in conjunction with the satellite.

\section{How Big? How Far? How Fast?}

Before directly addressing IRIS's scientific capabilities we must address
three important logistical questions: How big does IRIS have to be?  How far
away from the observer should it be?  How quickly will it traverse the night
sky?  These are related questions.  The farther away the satellite, the more
slowly it moves across the sky, but the smaller the solid angle it
subtends. Although we will address them in the context of the space-ground
configuration, similar questions will arise in the space-space mode.  We will
work, for now, in the limit of geometric optics, returning to the important
issue of diffraction below.

\subsection{Satellite Period}

To allow for as many observations at apogee as possible,
IRIS's orbital period should be at most a few days.
A period which is an integer number of days (or nearly so),
would allow  IRIS to be viewed by the same telescope at each apogee.
The importance, or desirability, of this last requirement 
will depend somewhat on how much special instrumentation and how large
a telescope is required to make optimal use of the satellite. 
We will take $r_a$ and $r_p$ such that $P\simeq 3\Day$.
For $r_p\simeq 6,900\km$, this gives $r_a\simeq 170,000\km$.

\subsection{Satellite Angular Velocity}

Just as the satellite must be close enough to subtend a minimum solid angle,
it must be far enough away to minimize its apparent angular velocity so one
can densely sample the light curve during an occultation.  For separation of
bright and dim images, such as in the occultation of a bright star to allow
the observation of an associated planet, the brightness contrast between the
sources is very large, typically $\gg 10^7$.  In the geometric optic limit,
the satellite completely blocks the bright source during occultations.
However, because of diffraction by the satellite, some of the light from the
bright source reaches the telescope (see the discussion in section
\ref{sec-diffract})\@.  Consequently, while one would like to search for
planets as close as possible to their associated stars, it is only for angular
separations of $\gtsim 0.2\as$ that the starlight is sufficiently dim not
to hide the planet.  Since the bright and dim sources are relatively well
separated, one can arrange for the satellite to occult only the bright source
(star) and leave the dim source (planet) unocculted.  The time over which one
can integrate the dim source is therefore approximately the crossing time of
the satellite for the bright source.  This must be long enough to allow the
planet to be detected with confidence above both the associated star and any
background.  (In the case of planet searches, this necessitates performing two
or more separate passes of the satellite to cover the complete area around the
star.)  For sources of comparable brightness one would like to be sensitive to
much smaller angular separations, $\phi \ltsim 1\mas$.  The relevant
integration time for establishing the compound nature of the source is
therefore the time it takes for the satellite to traverse the separation
between the two sources.

Naively, the angular velocity of the satellite is given by
$ \mu_{\rm proper}  = v_a/r_a $
where $v_a$ is the  velocity of the satellite at apogee,
and $r_a$ is again the apogee radius.
For a maximally eccentric orbit ($r_p=r_{p,\rm min}$, $r_a\gg r_p$)
\begin{equation}
\mu_p = \frac{v_a}{r_a} \simeq \frac{\sqrt{2GM_\oplus r_p}}{r_a^2}
= 375 \left(\frac{R_\oplus}{r_a}\right)^2 \as\second^{-1} .
\label{proper}
\end{equation}
Satisfying $\mu_p<1\as\second^{-1}$ would therefore require
$r_a > 1.2\times10^8\meter$.  
For imaging at the milli-arcsecond scale, 
this would allow only milli-second exposures,
confining one's attention to very bright source.

However, equation~(\ref{proper}) accounts only for IRIS's proper motion 
ignoring the apparent motion due to the Earth's rotation,
\begin{equation}
\delta\mu \simeq \frac{R_\oplus}{r_a}\frac{2\pi}{1 \Day}
= 15 \frac{R_\oplus}{r_a}\as\second^{-1} .
\end{equation}
(This is reduced by geometric factors having to do with 
the latitude of the telescope, the position of the source on the sky, \etc)
For $r_a\simeq 170,000\km$ in an eccentric orbit
(but $1,860,000\km$ for circular orbits),
the apparent motion due to the Earth's rotation matches IRIS's proper motion.
The satellite's apparent angular velocity at apogee can therefore
be reduced by putting it in a eccentric prograde orbit 
so that the velocity of the telescope due  to the Earth's rotation 
and the velocity of the orbiting satellite  
are equal at the time of occultation 
(or at least their components perpendicular to the line-of-sight are).  

\begin{figure}
\leavevmode\center{\epsfig{figure=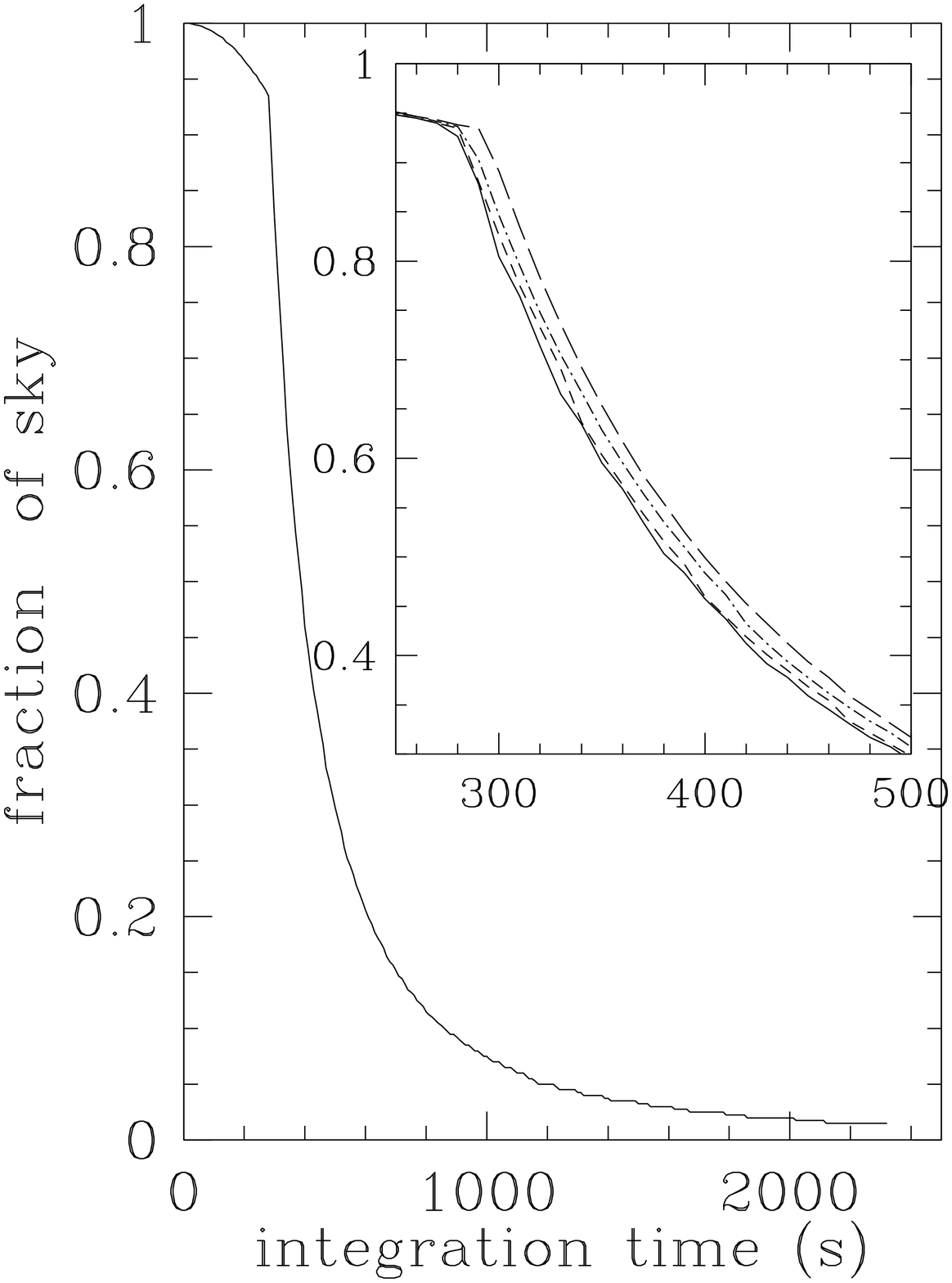,width=5in}}
\caption{Integration time as a function of latitude for
$r_p=6900\km$, $r_a=1.68\sci{5}\km$ and a period of 3 days.}
\label{fig:tint}
\end{figure}

How well does this velocity matching work?  For $r_a \simeq 1.7\times
10^8\meter$, it takes approximately $0.6\second$ for a satellite 300 meters
across to traverse its own length at apogee.  This would be the maximum
integration time in the absence of velocity matching.  In contrast, in
figure~\ref{fig:tint} we plot the fraction of the sky for which integration
times greater than or equal to a specific value can be achieved using a
$300\meter\times300\meter$ occulting satellite in an eccentric inclined orbit
from the Keck telescope with a 3 day period ($r_p = 6936\km$, $r_a =
1.69\times 10^8\meter$), and from the New Technology Telescope~(NTT), the Very
Large Telescope~(VLT) and Keck with a 4 day
period ($r_a+r_p=212,882\km$, for NTT $r_p=8901\km$; for VLT $r_p=9628\km$;
and for Keck, $r_p=10,278\km$).  The satellite is overhead at apogee.  In all
cases over $95\%$ of the sky integration times of 250--2300 seconds are
achieved.

The very significant increase in integration time due to velocity matching is
crucial to extend the reach of the satellite to the dimmest objects and the
smallest angular scales.  In a space-space configuration, one would attempt to
achieve these long integration times using station-keeping techniques. These
attempts are aided by the fact that orbital velocities about the L2 point are
quite low, typically $\ll 100 \meter\sec^{-1}$.

\subsection{Satellite Positioning}

Suppose we can reliably position IRIS in space
to a tolerance less than its linear dimensions.
The satellite must then be large enough to guarantee 
that scheduled occultations occur despite the uncertainty, $\Delta\delta$,
in the source location on the sky.
Typically $\Delta\delta\ltsim0.2\as\simeq 10^{-6}$.
Absolute positions can be determined better than this for bright sources,
but we may be interested in relatively faint and/or transient events,
for which such high precision astrometry is impracticable.
Moreover, much below $0.2\as$ the limiting technology 
is probably how well one can position the satellite, 
not how well one can decide where to position it.

Since we will schedule the most interesting occultations near apogee, 
where the  satellite's apparent  proper motion is a minimum,
we require that
\begin{equation}
r_a < x_\perp/\Delta\delta
\label{locate}
\end{equation}
Here we have resolved the satellite dimensions into components,
$x_\parallel$ and $x_\perp$,
parallel and perpendicular to the satellite velocity.
If $r_a\simeq1.5\times 10^8\meter$ and $\Delta \delta = 0.2\as$,
equation~(\ref{locate}) implies $x_\perp\gtsim 150\meter$.

The second effect influencing the choice for IRIS's size 
is that the fraction of the light  of an occulted source
that diffracts around the satellite decreases with the area of the satellite
(see section \ref{sec-diffract}).  
The higher the contrasts between bright and dim sources that one wants to study,
the larger IRIS needs to be.  
On the other hand, the larger IRIS is, the further away on the sky
the dim source must be from the bright source.
For high resolution imaging,  there is less constraint since in the Fresnel
limit the diffraction limit for the satellite at apogee 
is independent of the size of the satellite if
$\lambda < x^2/r \simeq 500\micron$.

We will use $x_{\parallel,\perp}=300\meter$ for IRIS in a ground-space
configuration, and $x_{\parallel,\perp}=100\meter$ for the space-space
configuration, though careful characterization of IRIS performance as a
function of size is essential. The optimum answer will depend on the exact
observing program and the nature of the objects to be occulted.

Once an occultation has successfully been executed,
one can use telemetry to pinpoint the absolute
position of the satellite to within about $0.5\meter$ (B.~Matisack, private
communication),  equivalent to $0.7 \mas$.  Relative positions could be
determined even more accurately.

\subsection{Satellite Mass}

The cost of launching a satellite increases with its mass.  
Hence there is some maximum allowable satellite area
$x_\perp x_\parallel$ which 
depends on the minimum film thickness one can tolerate,
the strength of any framework, etc\@.
Since the patches need to be opaque at the level of one part in 
$10^{4\hbox{--}6}$ they need be $10$--$15$ skin depths thick
in the relevant optical and IR wavebands.  
However, skin depths of good conductors in these bands are
much less than a micron, so 
by depositing  a thin coating of some good conductor (e.g.\ aluminum) 
on a substrate (e.g.\ mylar)
the minimum thickness is determined more by structural integrity 
than optical necessity---perhaps a few microns.
An aluminized mylar film $300\meter\times300\meter\times10\micron$ thick,
of which $3\micron$ is aluminum and the balance mylar
would have an approximate volume of $1\meter^3$ and an approximate mass of
$1600\kg$.

The rigidity of the satellite structure could be achieved 
by constructing IRIS from balloons inflated with low pressure gas.
Since the pressure in interplanetary space is so low, 
it would take little internal gas pressure to support such a structure.
Although micrometeoroids would puncture the film, 
the rate of gas leakage could be made acceptable,
especially if the balloon were  divided into small cells,
with no one single cell essential to the structural integrity of the whole.
Detailed calculations of the expected survival lifetime of the 
balloon are essential.
Additionally, a supporting structure could be deployed,
although the logistics of that could be complicated.
Holes totaling less than $10^{-(4\hbox{--}5)}$ of the area would
not materially affect optical performance.  
Moreover, only when the companion  holes in the front and back surfaces  
are aligned with the line of sight do they
affect IRIS's performance.
One difficulty may be ensuring that attitude and orbit corrections are
sufficiently gentle so as not to damage the framework or tear the film.
Also, one would like the amplitude of oscillations of the satellite to be small.
Detailed analysis of the modes of oscillation, damping times, \etc\
of a large balloon will be required, so too will determination
of the  maximum allowable amplitude of oscillation.

\section{Steerability: Space-Ground Configuration}

To change the satellite orbit to occult a particular object requires imparting
an impulse:
$\Delta{\vec p} = m\Delta{\vec v},$
resulting in a change of velocity
\be
\frac{\Delta\vec v_{\rm sat}}{v_{\rm sat}} 
= \frac{\Delta m_{\rm propellant}{\vec v}_{\rm ejection}}{m_{\rm sat}v_{\rm
sat}}
\ee
Some orbit correction can be done using the natural precession  of an inclined
eccentric orbit in the field of the Earth.
If $N$ is the number of major rocket-driven orbital corrections we would like
to make, then we must keep
$(\Delta m_{\rm propellant}/m_{\rm sat})\leq N^{-1}$,
where 
$\Delta m_{\rm propellant}$ is the typical mass of propellant expended per
orbit reconfiguration.
We therefore need  
\begin{equation}
v_{\rm ejection} \geq N \Delta{\vec v_{\rm sat}}
\end{equation}
For $r_a\simeq 1.68\times10^5\km$ we find $v_a \simeq 460\meter \second^{-1}$,
so if $\Delta{\vec v_{\rm sat}}\simeq v_{\rm sat}$, and $N \simeq 10^3$ this
implies $v_{\rm ejection}\gtsim 500 \kmps$.  $N={\cal O}(10^3)$ is a
reasonable number of corrections for an orbital period of $3\Day$, since it
implies a satellite lifetime of about 10 years, given one major correction per
orbit. The typical course correction should not however require 
$\Delta{\vec v_{\rm sat}}\simeq v_{\rm sat}$.
For example, to change the inclination of the orbit by $1^\circ$, 
requires  only
$\Delta{\vec v_{\rm sat}/ v_{\rm sat}} \simeq 0.015$.
So $10^{3-4}$ of these would require only
$v_{\rm ejection}\simeq 7\hbox{--}70 \kmps$.

If we wanted to use the satellite 
to observe transients like microlensing of LMC stars, 
then we would want to be  able to change the orientation
of the satellite orbit by $1$--$4^\circ$, the angular diameter of the LMC.  
Since there are only about 100 such events per year, 
we would be satisfied to be able to perform $10^3$ such course changes.
This requires only $v_{\rm ejection}\simeq 7\hbox{--}30 \kmps$.

The issue of steerability therefore comes down to the sizes of the course
corrections in which we are interested, the time scale over which we need to
move the satellite to the new orbit, and the state-of-the-art in high ejection
velocity drives (probably ion engines).  The smaller the corrections and the
longer we can wait, the less fuel we will need.  Since planets are not
transients, we can wait a relatively long time and use clever orbital dynamics
to do much of the work for us.  How many corrections we can make, therefore
depends on exactly how we use the satellite.  A reasonable program of
observations seems possible, though careful scheduling of observations and
clever design of orbital parameters will be essential.

Another intriguing possibility is to make use of the propellant-free
drive suggested by Drell, Foley and Ruderman~(1965).
In analyzing the orbital decay of the Echo satellite,
they observed that most of the decay could be attributed to 
the generation of Alfven waves, due to the motion
of this large conducting body across the magnetic field lines
in the Earth's  ionosphere.  They further noted that the drag can
be changed to a propulsion mechanism when a source of electrical
power is available on the satellite, and up to 50\% of the expended
power is available for pushing the satellite across the ambient field.
This mechanism could certainly be used for gross course  corrections
on IRIS, although fine corrections and attitude control
at the large apogee distances envisioned, would undoubtedly
have to be done by more traditional techniques.

Finally, since IRIS is essentially a $10^5\meter^2$ sail,
the possibility exists to harness the solar wind.  We have
not explored this further.

\section{Occultation Geometry: Seeing}

\begin{figure}
\leavevmode\center{\epsfig{figure=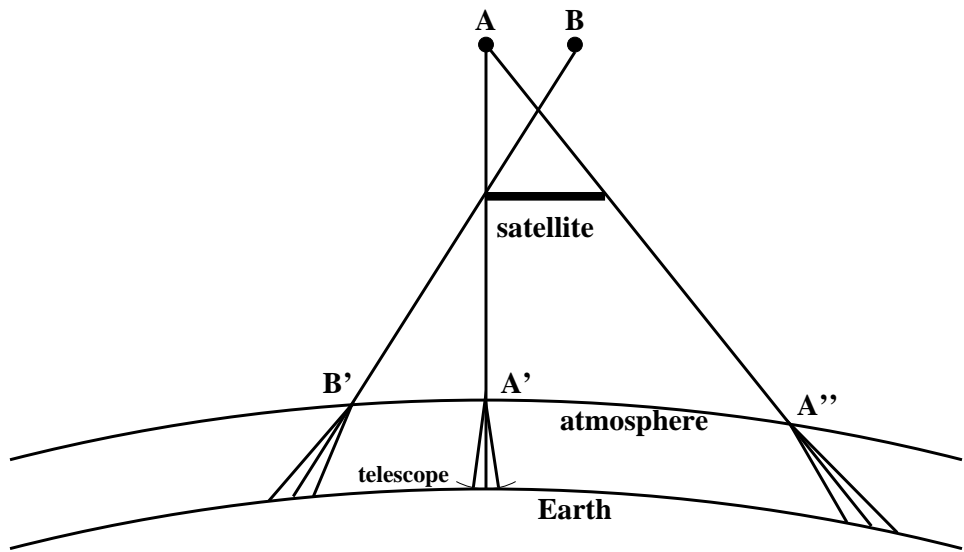,width=5in}}
\caption{Occultation of a binary source. Source A is occulted,
source B is about to be occulted. In the geometric optics limit,
rays from B strike the atmosphere at B' while
rays from A cannot strike the atmosphere between A' and A''.}
\label{fig:seeing}
\end{figure}

One might at first be concerned that atmospheric seeing will wash out the
angular resolution we wish to achieve.  Seeing would be a concern if we were
trying to achieve micro-arcsecond resolution.  In figure~\ref{fig:seeing}, we
illustrate the geometry when IRIS is occulting a binary source at the moment
when one source (A) is occulted and the second source (B) is about to be
occulted.  In the geometric optics limit, the light ray from B travels along a
straight line path from B past the satellite striking the atmosphere at $\rm
B'$. Rays from A cannot strike the atmosphere between $\rm A'$ and $\rm A''$
because they are blocked by the satellite.  The distance between $\rm A'$ and
$\rm B'$ is $\phi r_a$, and the angle subtended by $\rm A'B'$ is $\theta = r_a
\phi/ h_a$, where $h_a$ is some scale height of the atmosphere appropriate for
seeing.  For $r_a\simeq 1.6\times10^5\km$, and $h_a\simeq 10\km$,
$\theta>1\as$ for $\phi\gtsim0.6 \mas$.  Thus seeing does not spoil the
occultation for the $\phi\gtsim 0.1\mas$ of interest.

Nevertheless, seeing must be included when simulating IRIS images.  It is
particularly important in the detection of planets, because even the occulted
star is still much brighter than the planet.  Typically 99.97\% of the total
light is blocked in the B or V-bands, somewhat less at longer wavelengths.
Thus from the ground, one would like to use the best possible AO system
available.  The performance of the AO system can be aided by mounting a laser
on the bottom of the satellite and directing it at the observing telescope.
This would be better than a typical laser guidestar because the laser is above
the atmosphere so one could extract tilt and focus information.  A 10~W laser
with a $0.1^\circ$ beam is as bright as a $5\unit{mag}$ star, and is nearly
monochromatic.  The solar irradiation of the satellite is about 1.6~GW, more
than enough to power the laser.

In calculating the satellite's effects on images (see below), we include
seeing explicitly.  In the J and K-bands we use PSFs with diffraction limited
cores, and halos with FWHM of $1\as$ and an on-axis intensity of $10^{-3}$
of the on-axis intensity of the core.  These benchmarks are not far from
current capabilities and should be achieved in the years leading up to the
development of IRIS.

\section{Diffraction}
\label{sec-diffract}

Because we are interested in observing systems with very high contrast,
we must include the effects of diffraction.
The angular width of the satellite diffraction pattern 
in the region far from the satellite is 
\begin{equation}
\label{thetadif}
\Delta\theta_{\rm diff}\simeq \cases{
{\lambda\over x} & if ${\lambda\over x} \geq {x\over r_a}$ (Fraunhoffer limit),
\cr
&\label{diffraction}\cr
\left({\lambda\over r_a}\right)^{1/2} & if ${\lambda\over x} \leq {x\over r_a}$ 
(Fresnel limit).\cr}
\end{equation}
Here $x\simeq 300\meter$ is some typical size of the occulting patch.
For $\lambda\simeq 1\micron$ and $r_a\simeq 1.6\times10^8\meter$,
the transition from Fresnel to Fraunhoffer behavior occurs for 
$x\simeq \sqrt{\lambda r_a} \simeq  13\meter$. 
For $x=13\meter$ and $\lambda=1\micron$,
$\theta_{\rm diff}\simeq 8\times 10^{-8} = 16\mas$.

Inspection of (\ref{diffraction}), shows that once the size of the satellite
has reached $x_{\rm crit}\simeq\sqrt{\lambda r_a}$ no further improvement in
resolution can be obtained by increasing the size.  However, the fraction of
the intensity of the occulted source which is not blocked decreases with the
area of the occulter.  For this reason, we choose $x\gg 13\meter$.  The price
is that we must use the full Fresnel diffraction pattern of the sources as
occulted by the satellite and as observed through the telescope.

\subsection{Diffraction pattern of an occulting satellite}

The diffraction pattern for an arbitrary aperture ${\cal A}$ is given by
\be 
U = - \frac{Bi}{\lambda r's'} \int\!\!\int\nolimits_{\cal A} \tau(S) e^{ik
(r + s)}\,dS
\ee
where $B$ is a normalization constant related to the intensity
of the source, $dS$ is a surface element in the
aperture plane, $\tau(S)$ is the transmission function, $r$ ($s$) is
the distance between the observation (source) point and a point in the aperture
plane, and $r'$ ($s'$) is the distance between the observation (source) point
and the origin of the aperture plane.  Here we have assumed that the distances
between the aperture plane and the source and observation planes are large
and that all three planes are parallel.  We may expand $r+s$ in the
exponential in the limit of large separations.  However we must keep terms of
order $k\xi^2/2\Rsat > 1$ where $\xi$ is a dimension in the aperture plane.
The presence of this higher order term means that we are working in the
Fresnel limit of diffraction.

We begin by considering a rectangular hole in the aperture plane defined by
the transmission function
\be 
\tau (\xi,\eta) =  \cases{1, & $\xi\in[\xi_l,\xi_u]$,
$\eta\in[\eta_l, \eta_u]$ \cr 0, & otherwise}  .
\ee
The resulting diffraction pattern is
\be
\Urect (x, y)  =  \frac12 e^{i\alpha\left (\Delta x^2 + \Delta y^2\right)}
{\cal I} \left (x_l, x_u \right) {\cal I} \left (y_l, y_u \right),
\ee
where 
\be
x_{u,l} = \sqrt{\frac k{\pi\Rsat}} \left (\xi_{u,l} - \Delta x \right)
\quad{\rm and}\quad 
y_{u,l} = \sqrt{\frac k{\pi\Rsat}} \left (\eta_{u,l} - \Delta y \right).
\ee
Also
\be 
{\cal I} \left (x_l, x_u \right) = \left[ \fresnelC \left( x_u \right) -
\fresnelC \left( x_l \right) \right] - i \left[ \fresnelS \left( x_u \right) -
\fresnelS \left( x_l \right) \right],
\ee
where $\fresnelC$ and $\fresnelS$ are the Fresnel cosine and sine integrals,
If the entire
mask plane is empty ($\xi_l, \eta_l \rightarrow -\infty$, $\xi_u,\eta_u
\rightarrow \infty$) then 
$U_0 (x, y) = -i \exp[i\alpha\left (\Delta x^2 + \Delta y^2\right)]$.
Here $\alpha = k / 2\Rsat$, $\Delta x = x + \xsource$, 
$\Delta y = y + \ysource$,
and ($\xsource$, $\ysource$) is the position of the point source. 
Note that $U_0$ is normalized so that $\left| U_0 (x,y)\right| = 1$. The same
normalization is used for $\Urect$.
From Babinet's principle we can now easily construct the diffraction pattern
for a rectangular mask
\be
\Umask (x,y) = U_0 (x,y) - \Urect (x,y).
\ee

We now want to look at this pattern with a telescope which also causes
diffraction. The final pattern is given by an integral over the telescope
\be
U = \frac{\pi R^2}{\lambda R \sqrt{\pi}} \int_{-1}^1 dx \int_{-\sqrt{1-x^2}}^{\sqrt{1-x^2}}dy\, e^{-ikR (l_0 x + m_0 y)} \Umask (Rx, Ry).
\label{eqn:Uimag}
\ee
Here $R$ is the radius of the telescope and $(l_0, m_0)$ is the angular
position (in radians) of the observation point $(x_0, y_0)$ with
respect to the center of the telescope.  The denominator of the prefactor
normalizes the intensity in the observation such that 
\be 
\int_{-\infty}^\infty\int_{-\infty}^\infty dl_0\,dm_0 \left| U (l_0, m_0)
\right|^2 = 1.
\ee
We have treated the telescope as a circular aperture with a CCD at infinity
which is used to image the diffraction pattern.  

Once we have produced the image of the diffraction pattern we smooth it for
seeing.  In the J and K-bands we assume that AO systems will give us a
diffraction limited core with a Gaussian halo of $1\as$ FWHM\@.  The on-axis
intensity of the halo is taken to be $10^{-3}$ that of the core.  For the
B-band we assume the core is defined by a narrower Gaussian
(FWHM$\le0.25\as$) with the same on-axis intensity ratio.  As discussed in
the introduction, adaptive optics is well suited to this program since the
desired field of view for doing image separation is only about $4\as^2$.
The residual diffracted stellar images can be used as guide stars or, as
discussed above, a laser can be placed on the bottom of the satellite.
The details of the functional behavior of the PSF  (even the
rotational symmetry) are not as important as one's ability 
to characterize the PSF\@.  We assume the PSF can be characterized to 1\%\ in
the images presented below.

After smoothing, the images are formed by averaging over a standard wavelength
band and including a uniform background appropriate for this wavelength band
at a very good site such as Mauna Kea.  Finally we include counting statics on
the number of photons in each pixel via a Poisson distribution.

\section{The Satellite as a Source}

Just like a planet, IRIS would shine by both reflection and intrinsic emission.
A serious concern is that IRIS might appear brighter than the sources  
it is occulting.  

As we have already discussed, reflection of direct Sun-light,
Earth-light and even Moon-light will be an important concern.
Likely, it will be necessary 
to shield IRIS and to make it highly reflective.
By making IRIS reflective, 
one can choose which field  
to reflect into the field-of-view of the telescope.  
Because the field-of-view of interest is very small, 
finding a suitably dark field will be easy.
Keeping IRIS sufficiently stable that the same field is reflected 
into the telescope by all parts of the satellite for the entire exposure
is likely to be more difficult.
Alternately, one could allow (or even excite) 
small oscillations of the earth-facing surface
and thus smear the reflected fields over the full field-of-view.

\begin{figure}
\leavevmode\center\epsfig{figure=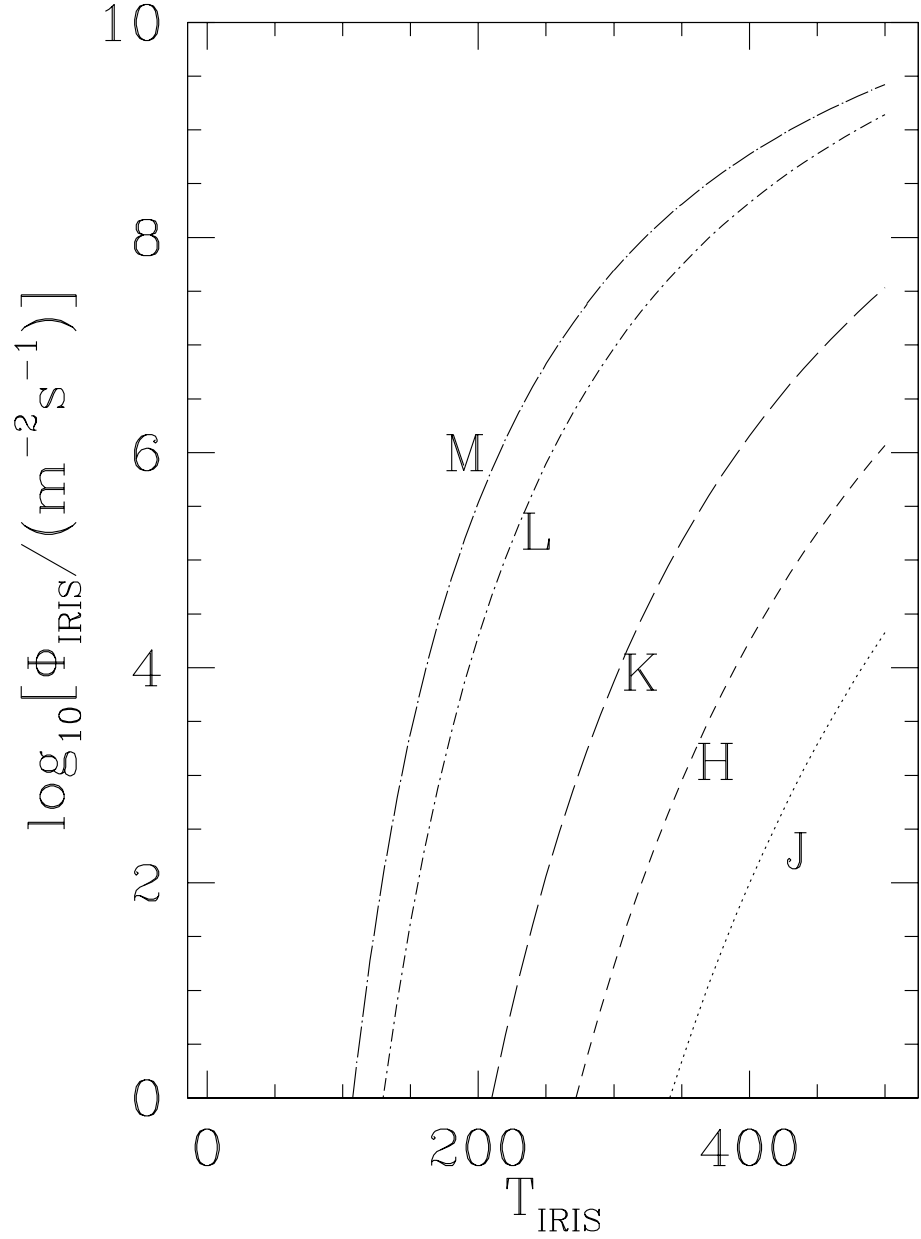, width=5in}
\caption{The flux from IRIS at apogee ($r_a=1.7\times10^5\km$),
as a function of the satellite's temperature  in the J, H, K, L and M-bands.}
\label{fig:irisflux}
\end{figure}

Intrinsic emission arises because IRIS is warmed by the sun.  
If allowed to come into thermal equilibrium as a blackbody,
IRIS's temperature would be comparable to the surface temperature of the Earth, 
\be T_{\rm IRIS} \simeq T_\odot 
\left({R_\odot\over r_\oplus}\right)^{1/2} 
\left({1 - \sigma_{\rm IRIS}\over 2}\right)^{1/4} 
\simeq  330\kelvin \left(1 - \sigma_{\rm IRIS}\right)^{1/4} .
\ee
Here $\sigma_{\rm IRIS}$ is IRIS's albedo.
The flux due to IRIS at  apogee is
therefore
\be
\Phi_{\rm IRIS} = {2 A\sigma T^4\over 2.7 k T 4\pi r_a^2}
\simeq 7.54\times10^8/\meter^2\sec \left({T\over 100\kelvin}\right)^3,
\label{irisflux}
\ee
where $A$ is the cross sectional area of IRIS\@.
This seems large, however, this is the flux integrated
over all wavebands. 
Because the peak of a black-body spectrum of temperature $100\kelvin$ is
at $\lambda\simeq50\micron$, 
and we are observing shortward of the K-band ($2.0$ to $2.4\micron$),
only a small fraction of IRIS's flux,
$\Phi_{\rm IRIS} f_\lambda(T)$,
is in the wavebands of interest.  Here
\be
f_\lambda(T) = \left. \int_{2\pi/\lambda_{\rm max}}^{2\pi/\lambda_{\rm min}}
{\omega^2 d\omega \over e^{\hbar\omega/k T_{\rm IRIS}} - 1} \right/
\int_0^\infty {\omega^2 d\omega \over e^{\hbar\omega/k T_{\rm IRIS}} - 1}.
\ee
For the K-band, this suppression is $f_\lambda = 1.9\times10^{-6}$ at
$T=330\kelvin$, 
and $f_\lambda = 1.4 \times 10^{-23}$ at $T=100\kelvin$.
In figure~\ref{fig:irisflux} we plot $\Phi_{\rm IRIS}f_\lambda$
for the various wavebands using equation (\ref{irisflux}).

We see that intrinsic emissions from the satellite 
are unimportant if the satellite is maintained below about $250\kelvin$.
Keeping IRIS that cool is merely a matter of keeping the
reflectivity around $90\%$.

\section{Planet Searches}

Planets shine in two ways, in reflected light and in emitted light.  In
reflected light, the brightness of a
Jupiter-like planet is $10^{-8}$, or less, times that of the star; falling off
as $1/r^2$ as the orbital radius increases.  In emitted light, a planet glows
as approximately a black body characterized by its surface temperature (though
molecular absorption can alter that dramatically in certain wavebands).  The
surface temperature is a strong function of the planet's age, the central
star's type and proximity, and atmospheric composition; however, typical
brightness ratios are still $\ltsim 10^{-8}$ except for very young gas giants
or planets very close to the central star (see e.g., Burrows \etal~1997).
Current state-of-the-art AO systems alone are not sufficient to enable
ground-based telescopes to to directly image planets.  Corresponding to these
two sources of luminosity, there are two wavelength regimes that are of
interest.

\begin{figure}
\leavevmode\center{\epsfig{figure=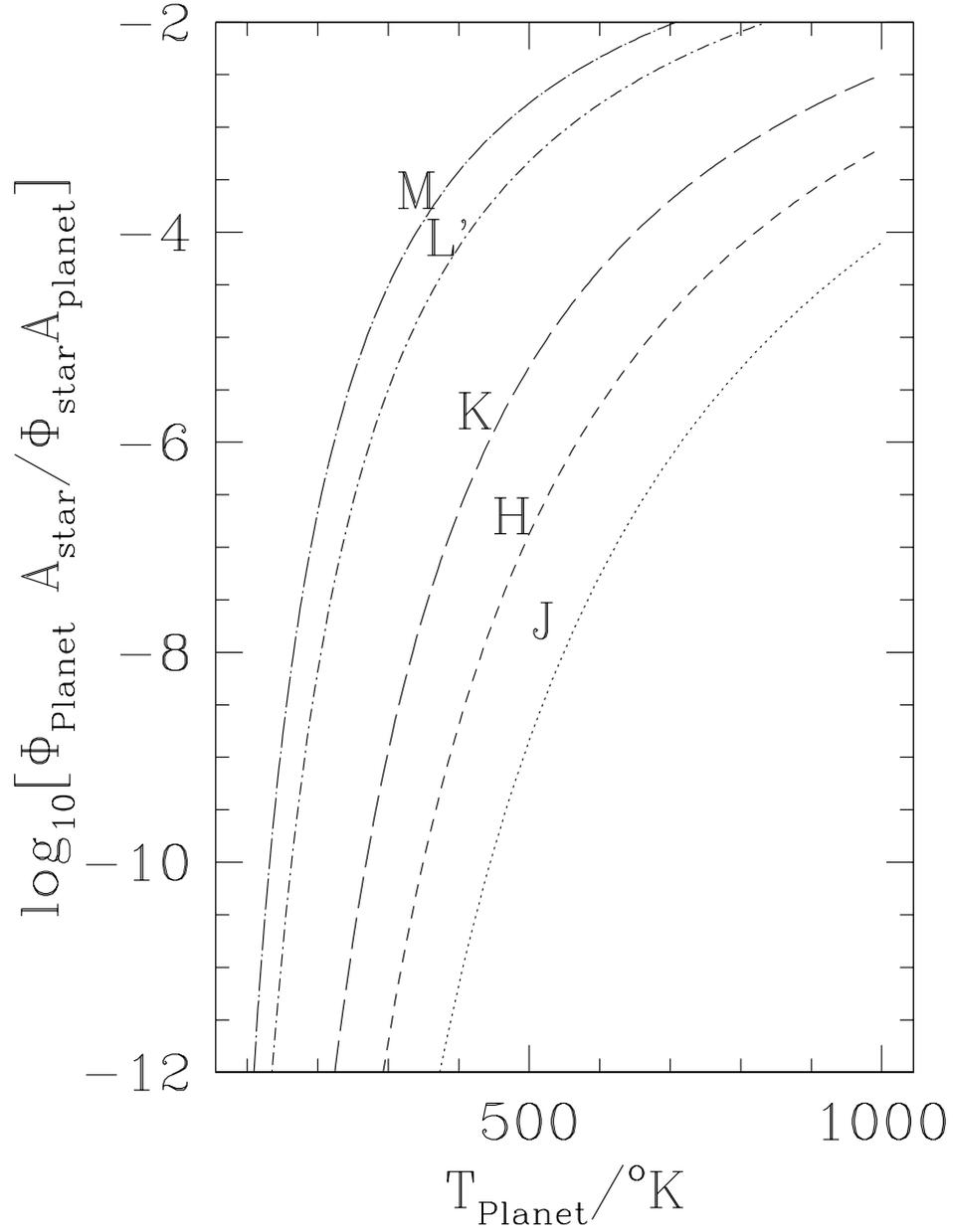,width=5in}}
\caption{Relative brightness of a Jupiter-sized planet and a G-dwarf star
($T=5788\kelvin$), as a function of the planet's surface temperature for
selected wave bands.
}
\label{fig:relbright}
\end{figure}

\begin{figure}
\leavevmode\center{\epsfig{figure=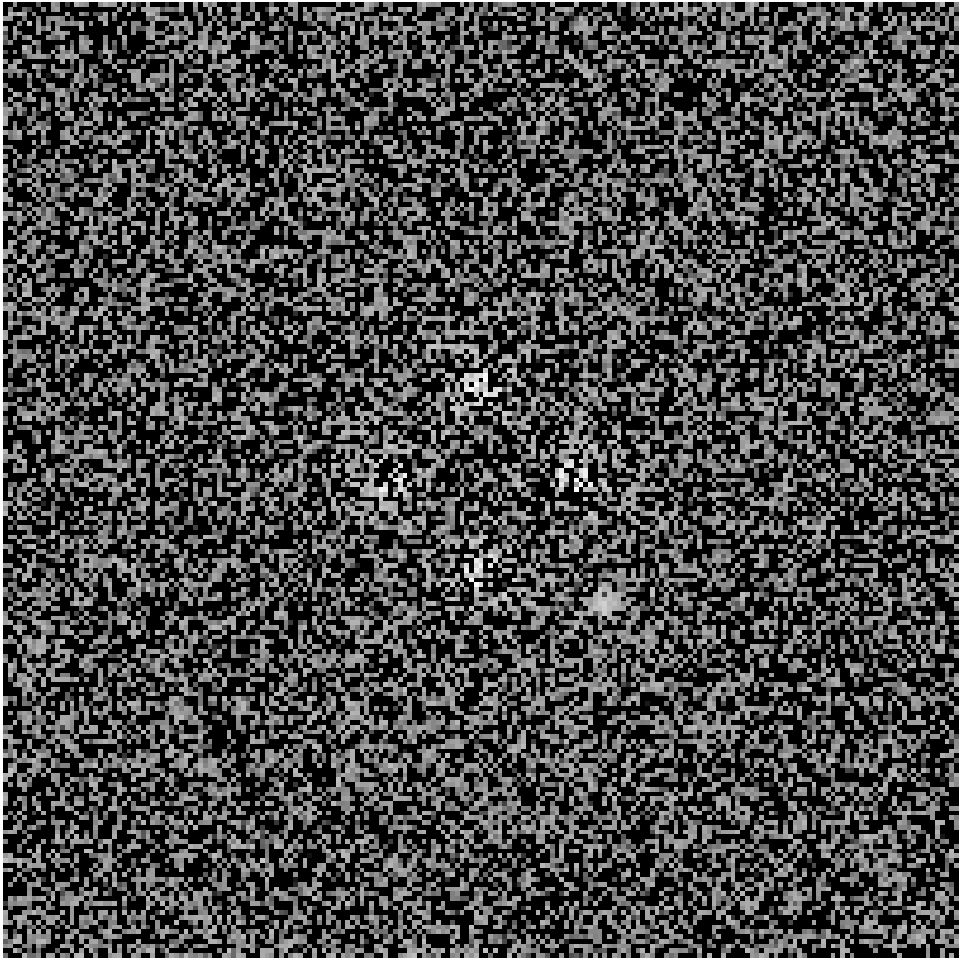,width=5in}}
\caption{The log of the intensity for a diffraction pattern in the K-band with
a planet $0.36\as$ from a star as seen in a $10$-m telescope.  The star is
$10\parsec$ away, has an apparent bolometric magnitude of 5, and is located at
the center of the satellite.  The planet is radiating as a blackbody at
$400\kelvin$\@.  We have assumed that the PSF can be characterized to 1\% and
have subtracted the PSF and a constant background.  The planet can be seen
along the lower right diagonal in the image.}
\label{fig:Kground}
\end{figure}

\begin{figure}
\leavevmode\center{\epsfig{figure=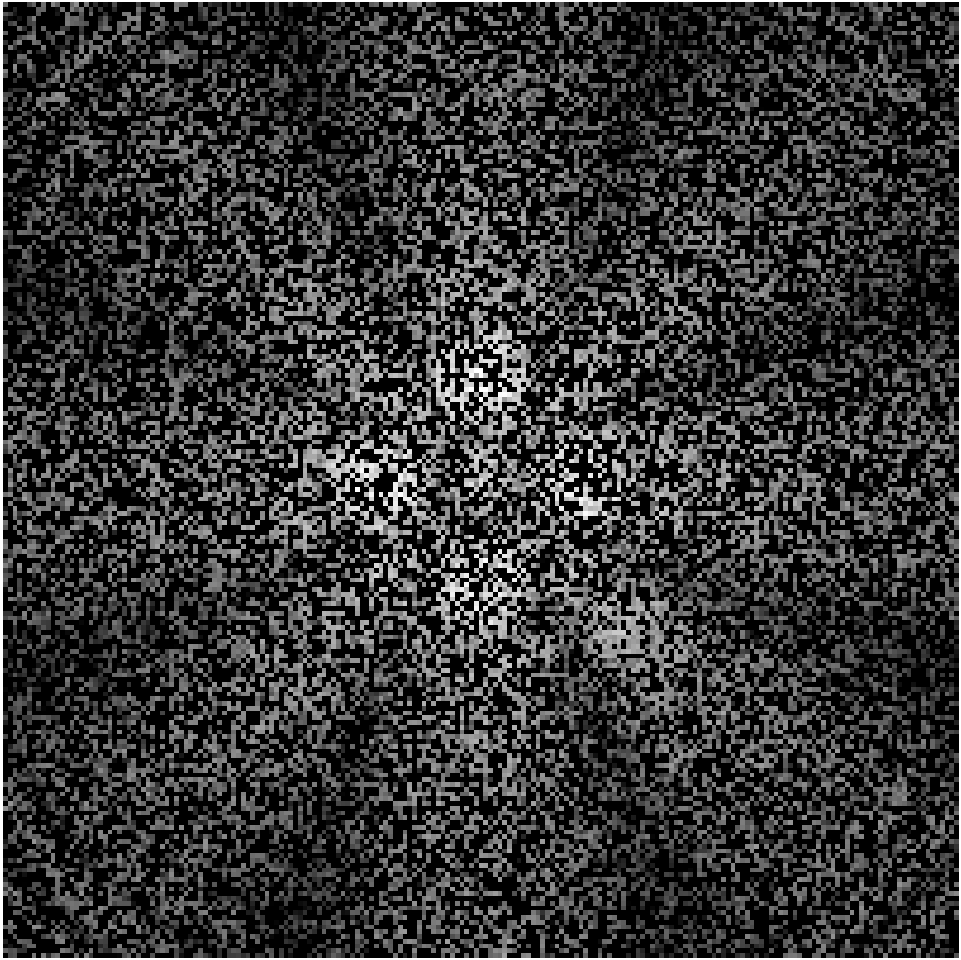,width=5in}}
\caption{The log of the intensity for a diffraction pattern in the K-band with
a planet $0.31\as$ from a star as seen in a $4$-m telescope in space.
The star is $10\parsec$ away, has an apparent bolometric magnitude of 5, and is
located at the center of the satellite.  The planet is radiating as a
blackbody at $400\kelvin$\@.  We have assumed a fully diffractive PSF and no
background.  Here 
the satellite is only $100\meter\times 100\meter$.  The planet can be seen
along the lower right diagonal in the image.}
\label{fig:Kspace}
\end{figure}

In the near IR ($1$--$10\micron$) the luminosity is
dominantly intrinsic emissions. 
In figure (\ref{fig:relbright}),
this is shown as a function of planet surface-temperature,
for a central star temperature of $5778\kelvin$. 
Although the planet-star contrast is smaller in the near IR,
the background is also much higher for ground-based telescopes.  
For K-band, the Mauna Kea
IR background is approximately 13.5~mag per square arcsecond.
Although longer wavelength bands are better in terms of planet 
intensity vs.\ stellar intensity, they are severely limited from the ground
by increased background.  
In figures~\ref{fig:Kground}-\ref{fig:Kspace}, 
we demonstrate the capabilities of the IRIS satellite in the near IR. 

In figure~\ref{fig:Kground}, we show a
Jovian planet with a surface temperature of $400\kelvin$,
orbiting at $3.6\AU$ from a 5~mag star $10\parsec$ from the Earth.
The images is a $1000\second$ exposure by a ground-based 10-m telescope.
The PSF has a diffraction-limited core 
and a $1\as$ (FWHM) Gaussian halo 
with an on-axis intensity $10^{-3}$ that of the diffractive core.
The image of the planet is clearly visible along the lower right diagonal. 

In figure~\ref{fig:Kspace}, we show the same system in a $1000\second$ exposure
by a space-based 4-m telescope with a purely diffractive PSF, but now with a
much smaller $100\meter\times100\meter$ satellite placed $56,000\km$ from the
telescope.  Here the planet is $3.1\AU$ from a 5~mag star $10\parsec$ from the
Earth.  Again the planet is clearly visible along the lower right diagonal.

\begin{figure}
\leavevmode\center{\epsfig{figure=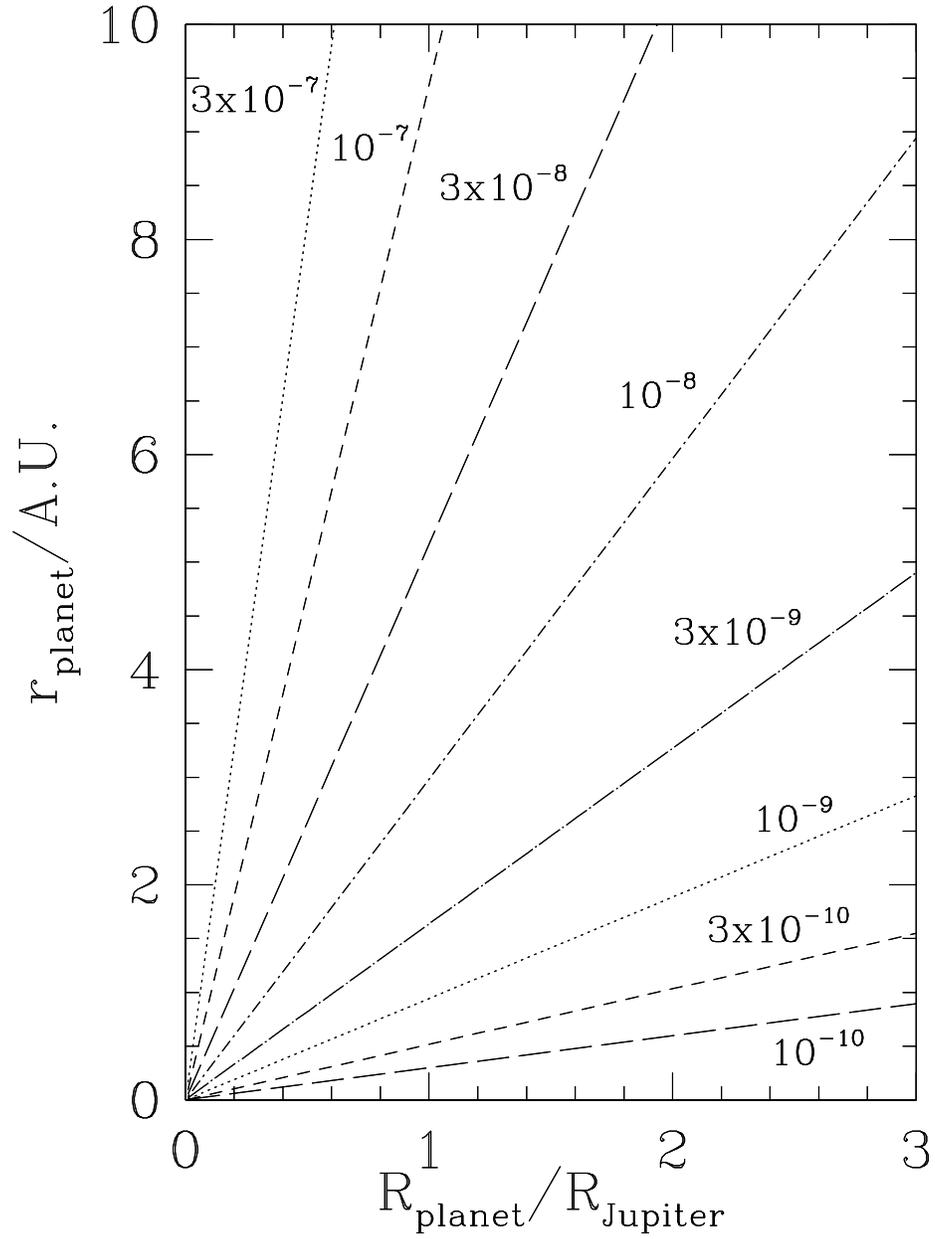,width=5in}}
\caption{Ratio of the reflected flux of a  planet to the flux
from the primary star, for an albedo of $0.5$ for orbital
separations of $0.2$--$10\AU$}
\label{fig:Vratio}
\end{figure}

At shorter wavelengths, the planets are best seen in reflection.
The typical visual albedos of gas giants are high, $\sigma\simeq0.5\pm 0.05$.
We take the ratio of flux from  the planet to flux from  the star to be
\be
{\Phi_P\over \Phi_\star} \simeq \sigma {\pi R_P^2\over  4\pi r_p^2}
\ee
where $R_P$ is the radius of the planet, and $r_p$ is its orbital radius.
For Jupiter orbiting at $1\AU$, this is $2.6\sci{-8}$.
In figure (\ref{fig:Vratio}) we plot 
$\Phi_P/\Phi_\star$ versus $R_P$ for orbital separations ranging
from $0.2\AU$ to $10\AU$.
These are easily related to the angular separation from the primary star
by 
\be
{\theta\over\as} = {R_P/{\AU}\over r_p/\parsec}
\ee

\begin{figure}
\leavevmode\center{\epsfig{figure=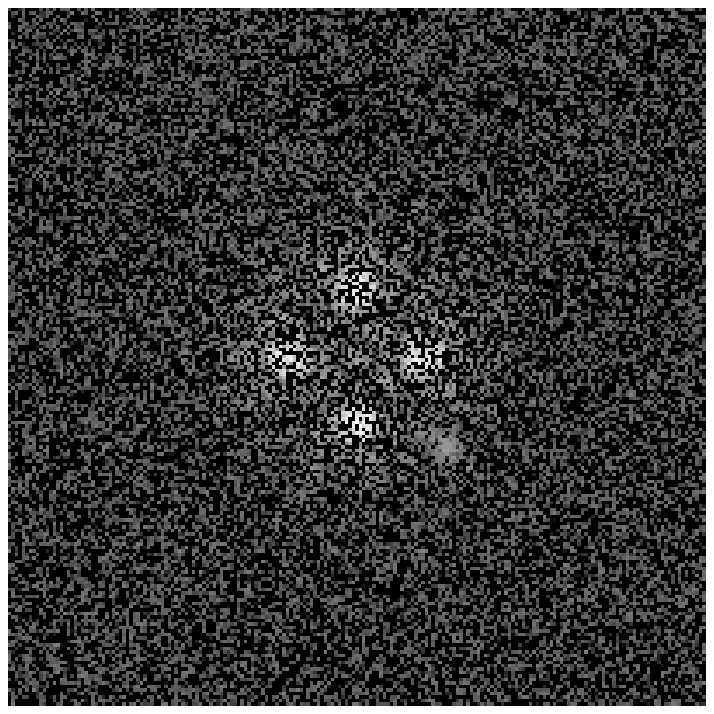,width=5in}}
\caption{The log of the intensity for a diffraction pattern in the J-band with
a planet 
$0.36\as$ from a star as seen in a $4$-m telescope.  The star
has an apparent bolometric magnitude of 5 and is located at the center of the
satellite.  The planet is reflecting light from the star with a relative
intensity of $3\sci{-8}$\@.  We have assumed that the PSF can be characterized
to 1\% and have subtracted the PSF and a constant background.  The planet can
be seen along the lower right diagonal in the image.}
\label{fig:Jground}
\end{figure}

In figure~\ref{fig:Jground}, we show a J-band image of a 
Jovian planet in reflected light. 
The brightness of the planet is $3\times10^{-8}$ of the central star,
and it is $0.36\as$ from the star. 
The image is a $1000\second$ exposure on a 4-m ground-based telescope,
using a PSF with a diffraction-limited core and 
a Gaussian halo with FWHM of $1\as$ 
and an on-axis brightness $10^{-3}$ that of the core.
The J-band background is taken to be 15.2~mag per square arcsecond.
Again the planet is clearly visible along the lower right diagonal.

\begin{figure}
\leavevmode\center{\epsfig{figure=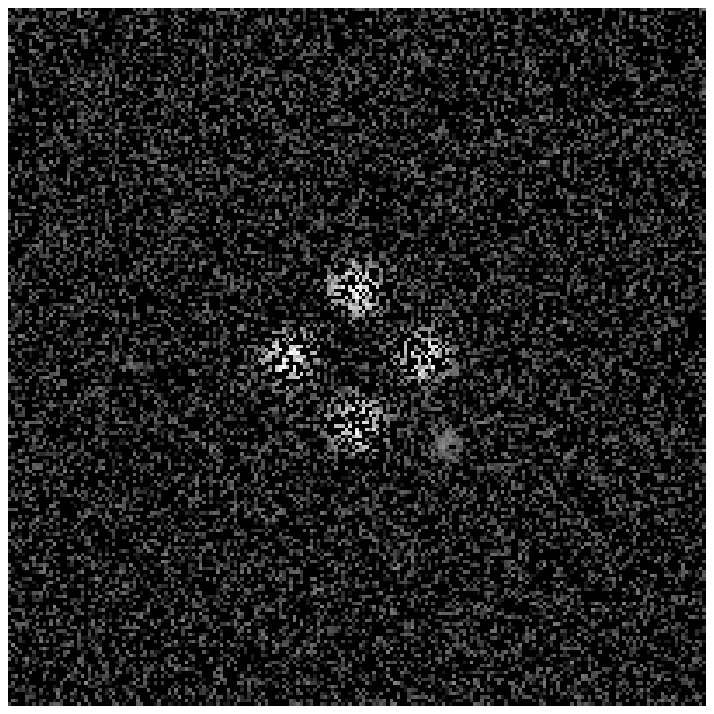,width=5in}}
\caption{The log of the intensity for a diffraction pattern in the B-band with
a planet $0.36\as$ from a star as seen in a $4$-m telescope.  The star
has an apparent bolometric magnitude of 5 and is located at the center of the
satellite.  The planet is reflecting light from the star with a relative
intensity of $5\sci{-8}$\@.  We have assumed that the PSF can be characterized
to 1\% and have subtracted the PSF and a constant background.  We have also
assumed seeing of $0.1\as$ (FWHM)\@. The planet can be seen along the
lower right diagonal in the image.}
\label{fig:Bground}
\end{figure}

In figure~\ref{fig:Bground}, we show a B-band image of a 
Jovian planet in reflected light, taken with 
a $1000\second$ exposure on a ground-based 4-m telescope.
The brightness of the planet is $5\times10^{-8}$ of the central star,
and it is $0.36\as$ from the star. 
The relevant backgrounds in the B-band is
$22\unit{mag} \as^{-2}\angstrom^{-1}$.
The PSF has a Gaussian core with a FWHM of $0.1\as$,
plus a Gaussian halo with FWHM of $1\as$ 
and an on-axis brightness $10^{-3}$ that of the core.
This core represents a factor of 2.5 improvement over the
seeing at the best sites.
Using a Gaussian core of $0.25\as$, this planet is not visible to the eye.
Image quality would also improve somewhat if we used
a 10-m telescope (we have not done so because of the much 
increased computer time necessary to produce the image),
but not enough to fully compensate for the seeing.
In fact, it would be better to deploy an AO system
which achieves a $0.1\as$ core on a 2 or 4-m telescope,
than a less effective AO system on a larger instrument.

\begin{figure}
\leavevmode\center{\epsfig{figure=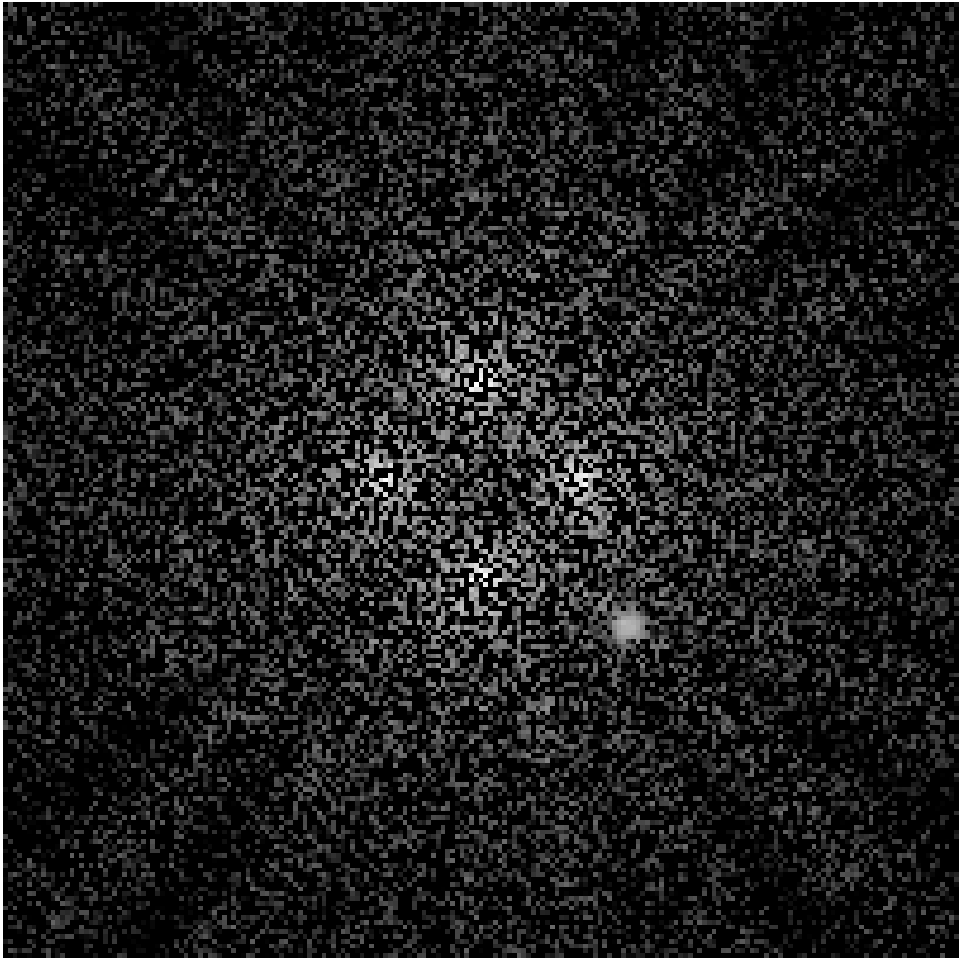,width=5in}}
\caption{The log of the intensity for a diffraction pattern in the B-band with
a planet $0.31\as$ from a star as seen in a $2$-m telescope in space.
The star has an apparent bolometric magnitude of 5 and is located at the
center of the satellite.  The planet is reflecting light from the star with a
relative intensity of $3\sci{-8}$\@.  We have assumed a fully diffractive PSF
and the same 
background as from the ground.  Here the satellite is only $100\meter\times
100\meter$.  The planet can be seen along the lower right diagonal in the
image.}
\label{fig:Bspace}
\end{figure}

In figure~\ref{fig:Bspace}, we show a B-band image of a 
Jovian planet in reflected light, taken with a $1000\second$ exposure
on a space-based 2-m telescope.
The brightness of the planet is $5\times10^{-8}$ of the central star,
and it is $0.31\as$ from the star. 
Once again, the satellite is only $100\meter\times100\meter$ and only
$56,000\km$ from the telescope.  A diffractive PSF, and a background of  
$22\unit{mag}\as^{-2}\angstrom^{-1}$ are used.

\begin{figure}
\leavevmode\center\rotate[r]{\epsfig{figure=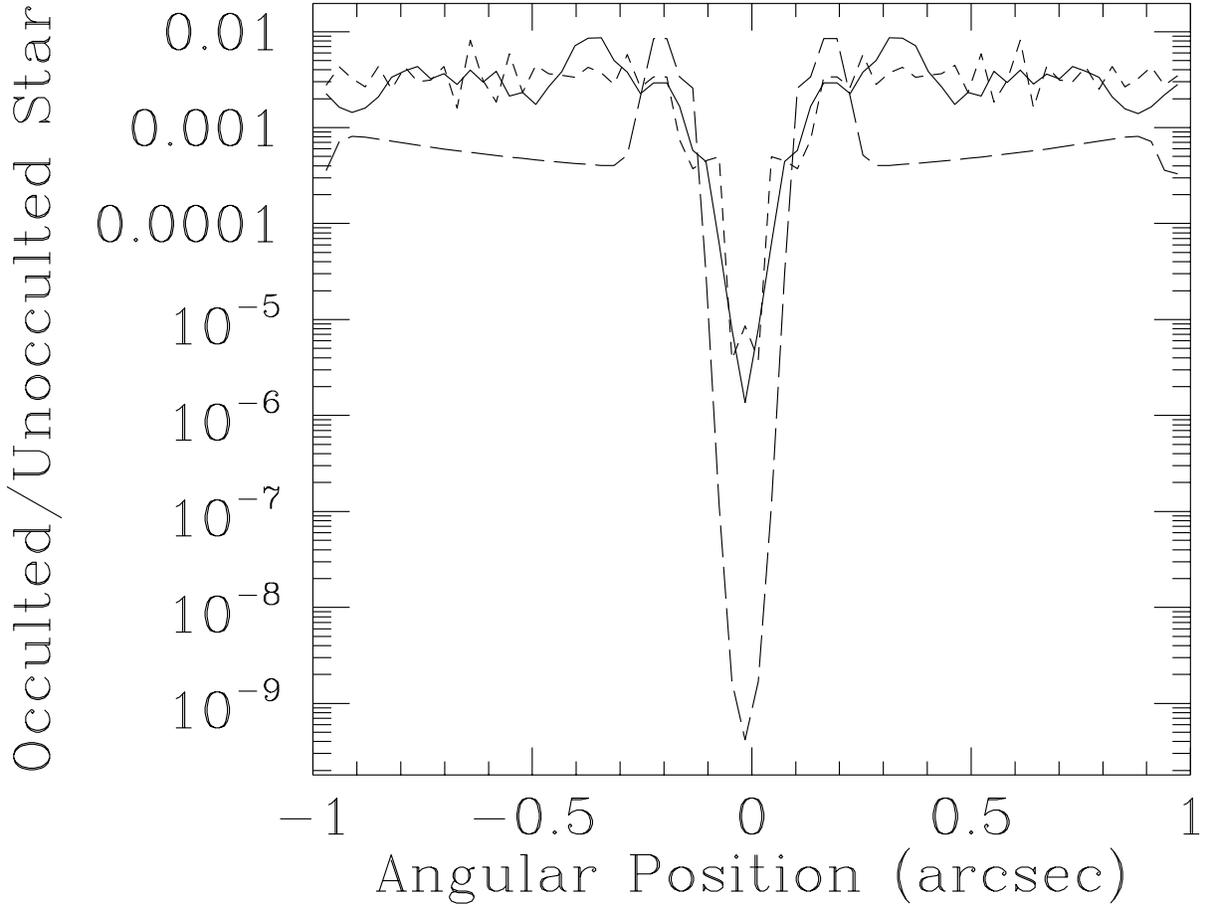,width=5in}}
\caption{Intensity ratio of an occulted and an unocculted star
($300\meter\times300\meter$ satellite at $168,000\km$)
as a function of angular separation from the star for a diagonal
slice across the field-of-view in B,J and K-bands. All images
have been convolved with a PSF consisting of a diffractive core in the J and
K-bands and a gaussian core in the B-band with a FWHM of $0.1\as$).
In both cases we include a gaussian halo (width of $1\as$ FWHM) with an on-axis
intensity of $10^{-3}$ of the core on-axis intensity.
The K-band  image (solid line) is for a 10-m telescope;
The B-band (long dashes) and J-band (short dashes) images are for a 4-m
telescope. No background or noise is included.
}
\label{fig:contrast}
\end{figure}

Finally, in figure \ref{fig:contrast} we plot the 
intensity ratios  
as viewed by ground-based telescopes
of a star occulted by  the satellite
and an unocculted star in the B, J, and K-bands
for a diagonal slice through the field of view  
averaged over $0.03\times0.03\as^2$ pixels.
(The diagonal slice being neither the most optimistic 
nor the most pessimistic.) 
This is arguably the true figure of merit for IRIS, 
as it  represents the extent
to which IRIS improves the contrast for the appropriate
observing  instrument.
In the K-band the observing instrument is  a 10-m telescope;
in the J and B-bands it is  a 4-m telescope.
In the J and K-bands we used a PSF with a diffraction limited core
and a Gaussian halo with a FWHM of $1\as$ and 
an on-axis intensity $10^{-3}$ that of the core.
In the B-band the core is a Gaussian with a FWHM of $0.1\as$.
The figure demonstrates that IRIS provides an improvement of
$(1\hbox{--}4)\times 10^{-3}$ in the intensity  in the wings for the J and
K-bands and $(4\hbox{--}8)\times 10^{-4}$ in the B-band.

\section{Image Resolution}

For sources of comparable brightness (such as individual images in a
gravitational microlensing event), one would like to be sensitive to much
smaller angular separations, $\phi \ltsim 1\mas$.  The satellite cannot be
reliably positioned so as to obscure only one source while transiting the
other.  Typically one source, then both sources will be occulted and then they
reappear in the same order.  The capabilities of IRIS to perform ultra-high
resolution imaging therefore rests on the ability to very accurately measure
the light curve of a compound source during an occultation event.  The
relevant integration time for establishing the compound nature of the source
is the time it takes for the satellite to traverse the separation
between the two sources on the sky.

\subsection{Binary Sources}
Consider two point sources  $A$ and $B$
separated by a small angle $\phi$,
with photon fluxes (${\rm photons}\meter^{-2}\second^{-1}$) 
$\Phi_A$ and $\Phi_B$ respectively 
in some appropriate waveband.
The rate at which photons from each source 
are collected by an instrument mounted on a telescope 
is $\Phi_iA_{\rm eff}$,
where $A_{\rm eff}$ is the telescope's collecting area 
times the detection efficiency.
If $\mu_{\rm app}$ is the apparent angular velocity of the satellite  
relative to the background stars,
then the expected number of  photons to arrive 
while the brighter of the two sources is eclipsed 
must be large enough to distinguish the dimmer source  
from either $0$ or $(\Phi_A+\Phi_B)$ at some satisfactory confidence level.
The number of photons we can collect while IRIS occults one source
and not the other is
\begin{equation}
N_\gamma = {\Phi_{\rm min}A_{\rm eff} \phi\over \mu_{\rm app} } 
\label{apparent}
\end{equation} 
where $\Phi_{\rm min}=\min(\Phi_A,\Phi_B)$.  This must be greater than some
minimum value $N_{\rm min}$.  The value of $N_{\rm min}$ depends on the
accuracy to which one wants to measure $\phi$, $\Phi_A$ and $\Phi_B$, and on
the background.

We have seen that typical occultations at apogee last $250-2300\second$
from a well-placed site (e.g.\ Mauna Kea) for $x_\parallel=x_\perp=300\meter$.
Since, for these values of $x_\parallel$, IRIS subtends
$0.36\as$, its average angular velocity across the field
of view is $0.16\hbox{--}1.5\mas\second^{-1}$.  Since the telescope is being
used essentially as a large light bucket in this mode, readout times $<1\sec$
for a CCD with $100\times100$ pixels is quite reasonable.

Values of $\Phi_i$ and $\phi$ vary depending on the target.
For gravitational microlensing the objects are 
stars in the Milky Way or in the LMC\@.  
An LMC target star (Alcock \etal~1996)
with apparent bolometric magnitude $m_b=18$, 
and effective temperature $T_{\rm eff}\simeq 5\times 10^3\kelvin$ has
$\Phi\simeq 8.5\times10^3\meter^{-2}\second^{-1}$.  
Since the dim image is typically not much dimmer than the unlensed star,
we will take this as $\Phi_{\rm min}$.
Typical angular separations of microlensed images are 
$\phi \simeq 1\mas$.
With $A_{\rm eff}\simeq 50\meter^2$,
equation (\ref{apparent}) gives a substantial
\be
N_\gamma \simeq (3\hbox{--}30)\times 10^5\phi\mas^{-1}.
\ee
This implies that $0.05\hbox{--}0.2\%$ changes in the flux
should be readily detectable.

Equation~(\ref{thetadif}) gives the diffraction angle, $\theta_{\rm diff}$,
as a function of wavelength and of satellite size and distance.
For B-band observations ($\lambda\simeq 4\times10^{-7}\meter$),
and $r_a\simeq 1.7\times10^8$,
if the two sources are closer together than about $10\mas$, 
then their diffraction patterns would overlap significantly.
The problem of reconstructing a binary source from the lightcurve
would then be complicated by diffraction-induced smoothing.  
In addition, the diffraction patterns must be averaged
over the spectrum of the source convolved with the transmission
function of the filter with which the objects are being observed.
This will degrade the resolution.  
Finally, the diffraction patterns must be integrated over the beam pattern
of the telescope, further degrading the resolution.
Including the  effects of both beam smearing and wavelength convolution,
Han \etal~(1996) have shown that for lunar occultation 
(which is also in the Fresnel limit) 
one can achieve resolution of approximately $3\mas$.
At first sight, 
since IRIS would be a half-way between the Moon and Earth
it would seem that a resolution of only $3\sqrt{2} \simeq 4\mas$
would be achievable for IRIS.  
However, the Moon traverses the sky at $0.54\as\second^{-1}$, a factor of
$300$--$3500$ times faster than IRIS's apparent velocity,
implying a proportional reduction in the number of collected photons,
and a consequent significant reduction in resolution.
Occultations with the satellite can also be repeated
for these smallest angular resolutions, 
thus enhancing the signal-to-noise.

As we have seen, 
once the size of the satellite 
has reached $x_{\rm crit}\simeq\sqrt{\lambda r_a}$
no further improvement in resolution can be obtained by increasing the size.
However, by opening several slits in the satellite we might narrow
the diffraction pattern and increase the resolution.  Moreover,
since we are able to determine the diffraction pattern a priori
very precisely (much more precisely than is possible for the Moon), 
we can model the light curve quite accurately and perform a fit to 
a binary source. 
This should help us to improve the resolution well below $10\mas$.

Furthermore, although we must still convolved the diffraction pattern as a
function of wavelength with the spectrum of the source, and with the
sensitivity of whatever detector and/or filters we may be using, we can avoid
somewhat the degradation in image quality that this would imply by doing
spectrometry.  In this fashion the diffraction peaks at different wavelengths
are not blended together, but rather can be used independently to reconstruct
the surface brightness of the source.  So long as the noise due to sky,
contaminating sources, or other backgrounds are small, this wavelength
spreading will not cause any decrease in signal.  Observations made at shorter
wavelengths could also help produce better resolution, though one must be
careful not to lose photon flux by going to wavelengths well below the
peak of the object's spectrum.

The integration over the beam function of the telescope will continue to
produce some degradation, if the peaks in the diffraction pattern of the
source are closer together than the diameter of the telescope.  Since $1\meter$
subtends $1.2\mas$ at $1.7\times 10^5\km$, it may be preferable to make one's
observations with a series of independent smaller telescopes ($1\meter$),
rather than one large diameter instrument.

\subsection{Compound Sources}

IRIS will be able to produce maps of compound sources, such as planets or
moons in our solar system, nearby stars and distant galaxies.  The resolution
of these images will be determined by the surface brightness of the source.
Given the photon count rates derived above, we estimate that for most systems
the resolution will be determined by requiring that the integrated flux over a
resolved patch correspond to an 18--20~mag point source.  We reserve further
discussion of all of these effects to future papers now in preparation.

\section{Conclusion}

The IRIS satellite in a space-ground configuration, will offer distinct
improvements for most currently envisaged ground based adaptive optics
approaches to planet detection.  The satellite blocks up to 99.97\% of the
light from a central star while leaving the balance in very well
characterized, very concentrated images.  Furthermore by doing the occulting
well above the atmosphere, the problems of scattering in the telescope are
minimized.  At the same time, IRIS would benefit from all efforts to improve
adaptive optics technology, particularly efforts to increase the fraction of
the PSF in the diffraction-limited core.  In the space-ground configuration we
have shown that planets can be detected approximately $0.35\as$ from a nearby
star in the B, J, and K-bands with only modest improvements in AO systems.
Any improvements will be immediately usable by the ground-based telescopes
making the observations and will improve the capabilities of IRIS\@.
Furthermore IRIS can be used in conjunction with other ground based detection
schemes such as coronographs and interferometers to improve upon them by
suppressing 99.97\% of the star's light.

In space-space configuration, a smaller IRIS satellite, coupled
with a 2--4-m class  space telescope with basic instrumentation
could greatly improve the prospects of seeing  colder planets, 
particularly in reflected light.  Such a configuration greatly improves the
B-band images due to the poor seeing from the ground and the difficulty of
building AO systems at these short wavelengths.  We have shown that with a
smaller satellite, closer to the telescope, we can also see planets
approximately $0.3\as$ from a star in the B and K-bands.  From space the
diffraction limit of the telescope, particularly in the K-band, becomes the
limiting factor.  In the space-space configuration station keeping may allow
longer integration times and the simpler orbital dynamics would reduce the
size and number of orbit corrections.

In the field of planetary physics, IRIS would allow us to search for new planets
around hundreds of nearby stars.  In the field of gravitational microlensing,
the measurement of the angular separations and orientations of the microlensed
images as a function of time would greatly increase what can be learned about
the physical properties of the lensing bodies.  This is especially true if one
could also obtain accurate measurements of the distance to the lens using some
paralactic method, or by noticing the departure of the light curve from
perfect time-reversal symmetry due to the Earth's motion.  In the field of
Galactic structure, we could greatly improve our understanding of disk
dynamics by obtaining parallaxes and proper motions of the nearby $1\kpc$, and
possibly study the core of the galaxy, as well as globular clusters at higher
resolution than has ever before been possible.  The prospects of image
separation with brightness contrasts of greater than $10^8$, of
milli-arcsecond resolution of binary sources, and of milli-arcsecond resolved
images of bright compound sources are all exciting.  The possibilities are
many.

We have shown that IRIS holds a great deal of promise but work remains to be
done.  More detailed PSF modeling, improved image identification algorithms,
and better use of wavelength information will aid in both image separation and
in high-resolution imaging observations.  Using planetary motion between
subsequent observations will aid in planet identification.  The incorporation
of more sophisticated models of planetary sources in both emission and
reflection is needed to more fully understand the size and temperature of
planets that can be observed.  Improved numerical techniques are required to
allow the simulation of observations by 10-m telescopes in the shorter
wavelength bands.  IRIS's capabilities must be more completely characterized
in the space-ground mode and in the space-space mode, for a variety of
telescope sizes.  This includes the improved exposure times for station
keeping and for a variety of satellite-telescope separations in the
space-space mode.  Finally an investigation of a varying transmission
coefficient over the surface of the satellite may lead to a truncated
diffraction pattern from the satellite thus increasing the suppression of the
light from the star and improving our ability to detect dim sources.

\acknowledgements
Allan Fetters did the detailed analysis of integration times
for the space-ground configuration and prepared the IRIS home page.  
David Rear contributed heavily to the programming and to analyses of 
imaging capabilities.
The authors would like particularly to thank 
C. Beichman for very extensive conversations on AO, planet finding,
IR astronomy and for encouragement and support;
Heather Morrison for educating them
about PSF's, filters, and all the nitty gritty astronomical details
theorists generally try to avoid;
and L. Close for telling them about state-of-the-art in AO.
They would also  like to thank
P. Taylor for assistance with questions on optics; 
W. Tobocman for special functions,
P. McGregor for information on IR backgrounds,
and A. de Laix, L. Krauss, H. Mathur and T. Vachaspati  for useful comments.
GDS would like to thank C. Alcock for 
very helpful discussions and encouragement
during the initial phase of this research, and
G. Marcy, B. Matisack, S. Drell and S. Tremaine for useful input.  
This work was supported by 
a CAREER grant to GDS from the National Science Foundation 
and by funds from CWRU.

\newcommand\ibid{\vrule height3pt depth-2.6pt width3em}

\clearpage

\def\color{}

\end{document}